\documentstyle[emulateapj]{article}
\begin{document}
\title{A Non-parametric Analysis of the CMB Power Spectrum}
\author{Christopher J. Miller and  Robert C. Nichol}
\affil{Department of Physics, Carnegie Mellon University, 5000 Forbes Avenue, Pittsburgh, PA 15213}
\author{Christopher Genovese and Larry Wasserman}
\affil{Department of Statistics, Carnegie Mellon University, 5000 Forbes Avenue, Pittsburgh, PA 15213}

\let\hat\widehat

\begin{abstract}

We examine Cosmic Microwave Background (CMB) temperature power spectra from
the BOOMERANG, MAXIMA, and DASI experiments.
We non-parametrically estimate the true power spectrum with no model assumptions.
This is a significant departure
from previous research which used either cosmological models or
some other parameterized form (e.g. parabolic fits).
Our non-parametric estimate is practically
indistinguishable from the best fit cosmological model,
thus lending independent support to the underlying physics
that governs these models.
We also generate a confidence set for the non-parametric fit and extract
confidence intervals for the numbers, locations, and heights 
of peaks and the successive peak-to-peak height ratios.
At the 95\%, 68\%, and 40\% confidence levels, we find functions that fit
the data with
one, two, and three peaks respectively
($ 0 \le \ell \le 1100$).
Therefore, the current data prefer two peaks
at the $1\sigma$ level. However, we also rule out a constant temperature
function at the $> 8 \sigma$ level.
If we assume that there are three peaks in the data, we find
their locations to be within $\ell_1$ = (118,300), $\ell_2$ = (377,650),
and $\ell_3$ = (597,900). We find the ratio of the first peak-height
to the second $(\frac{\Delta T_1}{\Delta T_2})^2 = (1.06, 4.27) $
and the second to the third  $(\frac{\Delta T_2}{\Delta T_3})^2 = (0.41, 2.5)$.
All measurements are for 95\% confidence.
If the standard errors on the temperature measurements were
reduced to a third of what they are currently,
as we expect to be achieved
by the MAP and Planck CMB experiments, 
we could eliminate two-peak models at the 95\% confidence limit. 
The non-parametric methodology discussed in this paper has many 
astrophysical applications.

\end{abstract}

\section{Introduction}

There has been growing evidence for the existence of peaks
and valleys in the temperature power spectrum of the CMB.
\relax From a theoretical standpoint, 
such features are a direct
result of the physics in the primordial photon-electron plasma,
predicted by gravitational instability models of
structure formation (Peebles \& Yu 1970; Hu and Sugiyama 1995).  
These features are important for constraining the cosmology of our
Universe. For instance, in many models, the ratio of the height of the first peak
to the second peak is dependent on the spectral tilt, $n_s$ and
the baryon fraction, $\Omega_{baryons}/\Omega_{matter}$.
The ratio of the third peak to the
second peak is dependent on $\Omega_{matter}h^2$ and $n_s$
(see Hu et al. 2001 for further discussion).

Most often in the literature, the CMB power spectra are fit to 
a suite of cosmological models (Tegmark et al., 1999,2000; Jaffe et al. 2001).
These physical models are well-motivated and sophisticated,
but they contain many free parameters 
(\emph{e.g.}, eleven in the work of Wang, Tegmark, and Zaldarriaga 2001-- WTZ),
some of which are 
unknown (ionization depth, contribution from gravity waves) 
or degenerate (e.g. see Efstathiou 2001). Typically, some sort of likelihood analysis is
performed to determine which cosmological model best fits the data.

There is however, another approach:
place constraints on the features
of the power spectrum and use these features to determine the cosmological parameters.
The assumptions here are that the peaks and valleys are best described
by the broad range of cosmological models (as in Hu et al.)
or by parabolas or some other chosen function 
(as in  Knox \& Page 2000 and de Bernardis et al. 2001).
A potential problem in all of these approaches
is that it is difficult to get valid statistical confidence intervals
(see Section 2).
There is also the concern that the fitted features
may be artifacts from the multitude of assumptions.

In this paper, we take what may be considered a more conservative
approach: we make no assumptions whatsoever about the true underlying function.
Our new statistical technique is non-parametric 
and allows for valid confidence intervals to be measured for peak characteristics.
One theme of our work is that confidence intervals for any quantity
of interest can be extracted from a confidence set for the unknown spectrum.
These techniques for fitting and inference
are applicable to a wide variety of
astrophysical data-analysis problems.

\section{Overview of Nonparametric Analysis}

In general,
non-parametric statistical methods estimate
functions without imposing a finite-dimensional
parametric form.
The resulting estimates are obtained 
by carefully smoothing the data to balance bias and variance.
See Hastie and Tibshirani (1990) for details and examples.

The CMB data, after suitable preprocessing, take the form
$(X_1, Y_1), \ldots, (X_n,Y_n)$
where $X_i$ is the multipole moment (usually denoted $\ell$) ordered
according to increasing $X$, and
$Y_i$ is the estimated power spectrum at $X_i$ (usually denoted
$C_{\ell}$ with some constants).
Let $f(X_i)$ be the true power spectrum at $X_i$.
Then,
\begin{equation}\label{eq::model}
Y_i = f(X_i) + \epsilon_i
\end{equation}
where
$\epsilon_i$ is the error in estimating $f(X_i)$.
We require that the
$\epsilon_i$ are uncorrelated, zero-mean Gaussians.
Therefore, we take the uncorrelated statistical errors as
given by the experiments. There are additional, correlated
noise terms in all of the experiments due to calibration,
beam width and pointing uncertainties, which were not 
used in our analysis. The magnitude of these correlated errors
is typically 5\% - 20\%, although the Boomerang effective
beam uncertainties can be higher for $\ell > 600$. The method
we use in this work can be generalized to include correlated
errors (Genovese et al., in preparation).
We assume that $f^2$ is integrable;
otherwise, we make no further assumptions.
Our non-parametric technique 
yields a vector $\hat{\bf f}$
that estimates the vector ${\bf f} = (f(X_1), \ldots, f(X_n))$
of the true spectrum's values at the $X_i$s (see Figure 1 top).

After we perform the fit,
we need to quantify the uncertainty to make any inferences.
We begin by constructing a set of vectors $C_n$ from the data
that traps the true power spectrum,
$\bf f$, with probability $1 - \alpha$,
where $1 - \alpha$ is a pre-specified confidence level.
With $C_n$ we can derive a confidence interval for any quantity of interest.
Consider, for example, the number of peaks.
For any vector ${\bf f}\in C_n$, define $P({\bf f})$ to be the
number of peaks in ${\bf f}$.
Then, because $C_n$ contains the true spectrum with confidence $1 - \alpha$,
the range on the measured peaks is 
$\left(\min_{{\bf f}\in C_n} P({\bf f}),\max_{{\bf f}\in C_n} P({\bf f})\right)$.
We can find confidence intervals for the locations and heights of peaks
in a similar way. There are important advantages for these confidence
measurements over ``standard'' $\chi^2$ or even Bayesian techniques,
which we discuss further in the next section.

\section{Methodology}

Refer to the model in equation (\ref{eq::model}).
Let the functions
$\phi_1, \phi_2, \ldots$ be an orthonormal basis
over the range of the observed $X_i$s.
The choice of basis is somewhat important. For instance, if we were
fitting to a galaxy spectrum, with highly peaked emission lines
on top of a smooth, broad continuum, a wavelet basis would
allow for the simultaneous fitting of broad and narrow features.
On the other hand, 
a cosine basis would require large amplitude high frequency terms
to match the
emission lines. This might cause the continuum fit to be wiggly.
In our work, we use a discrete cosine basis, $\phi_1 (x) \equiv 1$,
$\phi_2 (x) = \sqrt{2} cos(\pi x)$,
$\phi_3 (x) = \sqrt{2} cos(2\pi x), \ldots$,
since it has well
determined properties for confidence limits.
For simplicity in the derivation, 
we assume that
the variance (second moment about the mean) of each $\epsilon_i$ is the same
value $\sigma^2$.
This is not necessary in practice, nor was it assumed in the full analysis.

Any square integrable function $f$ can be expanded as
$f(x) =\sum_{j=1}^\infty \beta_j \phi_j(x)$.
Estimating $f$ amounts to estimating the $\beta_j$'s.
If we have chosen a good basis for representing $f$,
the higher-order terms in this series will tend to decay rapidly.
Hence, 
we can approximate the infinite sum with finite sum
$f(x) \approx \sum_{j=1}^N \beta_j \phi_j(x)$.
So, to estimate the underlying function, we need to
find the $\beta$s.

Let $Z_j = N^{-1/2}\sum_{i=1}^N Y_i \phi_j (X_i)$,
for $j=1,\ldots, N$. Based on the theory in Beran (2000), we 
take $N = n$. This choice ensures that the estimate of $f$ 
is optimal and that the resulting confidence intervals remain
valid.
It can be shown that each statistic
$Z_j$ has approximately a Gaussian distribution
with mean $\theta_j = \sqrt{n}\beta_j$ and variance $\sigma^2$.
This re-parameterization means that we need to estimate
$\theta=(\theta_1, \ldots, \theta_n)$.
Given an estimate $\hat\theta$ of $\theta$,
we can estimate $f(X_i)$ via
$\hat{f}(X_i) = (1/\sqrt{n})\sum_{j=1}^n \hat{\theta}_j \phi_j(X_i)$.

We could use $Z_j$ as an estimate of $\theta_j$,
but this yields a poor estimate of $f$ 
because it is too variable.
A smoother estimate can be obtained by damping the higher frequency terms in the expansion.
In statistics, this is called a ``shrinkage estimator''.
We consider shrinkage estimators of the form
$\hat{\theta}= (\gamma_1\,Z_1, \gamma_2\,Z_2, \cdots, \gamma_n\,Z_n)$,
where $1 \ge \gamma_1 \ge \gamma_2 \ge \cdots \ge \gamma_n \ge 0$
are called the shrinkage coefficients.
The smaller $\gamma_j$, the smaller the contribution of $\phi_j$ in the estimating
expansion for $f$.
With the cosine basis, for example,
such shrinkage estimators damp down the contribution from high-frequency terms.

Every choice of 
$\gamma =(\gamma_1, \ldots, \gamma_n)$
gives an estimate $\hat{\theta}^\gamma$
which then yields an estimate
$\hat{f}^\gamma$ of the function $f$.
We would like to choose
$\gamma$ to minimize the mean squared error,
${\rm MSE}(\gamma) =\langle \int (f(x) -\hat{f}^\gamma(x))^2 dx\rangle$.
Unfortunately ${\rm MSE}(\gamma)$ is unknown, because it depends on the true $f$,
but it can be estimated by Stein's Unbiased Risk Estimator (Stein 1981):
\begin{equation}
\hat{{\rm MSE}}(\gamma) 
  = \sum_j \left[\sigma^2 \gamma^2_j + \left(Z_j^2 - \sigma^2\right)(1-\gamma_j)^2\right].
\end{equation}
We use the Pooled Adjacent Values (PAV) algorithm 
(Robertson, Wright, and  Dykstra, 1988)
to minimize $\hat{{\rm MSE}}(\gamma)$
as a function of $\gamma$
while maintaining the ordering constraint on $\gamma$.
The minimizer is denoted $\hat{\gamma}$ and the final estimate
is therefore $\hat{\theta} = (\hat{\gamma}_1\,Z_1,\hat{\gamma}_2\,Z_2,\cdots,\hat{\gamma}_n\,Z_n)$.

Next we construct a confidence set, $C_n$ for the vector of
function values at the observed data,
${\bf f}_n = (f(X_1), f(X_2), \cdots, f(X_n))$. 
Throughout the paper
we have said that the confidence sets are ``valid.''
Formally, what this means is that
for any $c > 0$,
\begin{equation}
\lim\footnote{Strictly speaking, we should use a $\limsup$, which refers to the
largest difference between the true coverage and the claimed coverage that can
be obtained as the sample size gets large}_{n \rightarrow \infty}~ \sup_{||{\bf f}_n|| \le c} 
\left|{\rm Pr}({\bf f}_n \in C_n) - (1-\alpha)\right|\rightarrow 0
\end{equation}
as $n\rightarrow \infty$,
where the operation $||a||$ denotes
$\sqrt{n^{-1}\sum_i a_i^2}$.
This means that for large $n$, the confidence set traps
the values of the true function with probability very close
to $1 - \alpha$.
The confidence set is
an ellipse of the form
\begin{equation}
\label{eq:radius}
C_n =\left\{ \theta :\ n^{-1}\sum_j (\theta_j - \hat{\theta}_j)^2 \leq
\hat{MSE}(\hat{\gamma}) + n^{-1/2}\tau  z_\alpha \right\}
\end{equation}
where $z_\alpha$ is the number that has probability
$\alpha$ to the right under a standard Gaussian.
For instance, if we choose 95\% confidence ($\alpha = 0.05$), then $z_\alpha = 1.645$, while
for 67\% confidence ($\alpha = 0.33$), the confidence ``radius'' is smaller with $z_\alpha = 0.44$.
Here, $\tau$ is defined similar to Beran (2000):
\begin{equation}
\tau^2 = 4\,\sigma^2\,\sum_j (Z^2_j - \sigma^2)(1 - \hat\gamma_j)^2
 + 2\,\sigma^4\,\sum_j (2\hat\gamma_j -1)^2.
\end{equation}

We can in turn write the confidence set for ${\bf f}$ as
\begin{equation}
{\cal D}_n = \left\{ {\bf f} :\ {\bf f} = n^{-1/2}\,\sum_{j=1}^n \theta_j \phi_j \mbox{ and } \theta \in {\cal C}_n \right\}.
\end{equation}
Here, the true power spectrum ${\bf f} = (f(X_1),...,f(X_n))$ is 
estimated at each of the original data points.
Prior information about $\bf f$ of the form ${\bf f}\in{\cal P}_n$
allows us to replace ${\cal D}_n$ with
${\cal D}_n \cap {\cal P}_n$
while maintaining a $1 - \alpha$ confidence level.
In particular, 
we take ${\cal P}_n$ to be the set of vectors ${\bf f}$
corresponding to spectra with zero to three peaks for $\ell \le 1100$.
\relax From the confidence set ${\cal D}_n \cap {\cal P}_n$,
we can derive confidence sets for any interesting 
feature of ${\bf f}$.

A key point is that the resulting intervals on any measured
quantity are \emph{simultaneously valid},
meaning that all of the intervals contain the corresponding true quantity
with probability $1 - \alpha$. 
In contrast, deriving $1 - \alpha$ confidence intervals from a collection
of individual chi-squares does not obtain $1 - \alpha$ simultaneous coverage,
but often substantially lower coverage. For example, a common technique to
determine the 95\% confidence range for a specific parameter is to ``marginalize''
over the other parameters (see Tegmark et al. 1999,2000).
Such a technique will provide full coverage for that one parameter. However,
when parameter ranges are combined, the confidence is lower than 95\%.
Bayesian intervals derived from a posterior distribution suffer from a similar
problem in that the long-run frequency that the interval contains the true quantity
may be much less than $1 - \alpha$.

\section{Results}
Figure 1 shows the combined data
from BOOMERANG, MAXIMA, and DASI
(Halverson et al. 2001; Lee et al. 2001; and Netterfield et al. 2001).
The bottom panel compares our non-parametric fit to the
the fit of WTZ who use more experiments than the three examined here.
Recall, our fit requires no assumptions about the data
or the underlying cosmology. The WTZ fit, on the other hand,
requires an 11 dimensional parameter space, with numerous
prior assumptions placed on those parameters (for example, the
Hubble constant is constrained to $72 \pm{8}$ km s$^{-1}$Mpc$^{-1}$).
The agreement between the two fits is very good,
considering the difference in methodology.

The power of this technique
lies in the ability to make quantitative statements about the true function 
with some specified confidence.
First, we checked the 95\% confidence set and find
that every function within this set has at least one peak. Specifically,
we set $\alpha = 0.05$ and determined the ``radius'' of this
confidence ellipse (ie. the right side of the inequality in Eq \ref{eq:radius}).
We then searched all possible functions with zero
to three peaks over the specified range in $\ell$ to see if the condition in
Eq \ref{eq:radius} was met. Our definition of one peak requires the sorted data
(according to increasing $\ell$) to have a section with increasing temperature
followed by a section with decreasing temperature.  For $\alpha = 0.05$, 
no zero-peaked functions met this condition. Our set of zero-peaked functions includes
those with constant temperature as well as those with either increasing temperature
or decreasing temperature, but not both.
We rule out functional forms with zero peaks at the level 95\%
confidence level. We also compared specifically against the best fit constant function
for the power spectrum. We rule out a flat temperature spectrum (based on the weighted
average of the data) at $> 8\sigma$ confidence.  We perform the same analysis at the 68\% confidence
level (e.g. $\alpha  = 0.32$ in Eq \ref{eq:radius}).
The 68\% confidence set rules out all single-peak functions. So at
the one sigma level, we have found at least two peaks in the data.
Only in the 40\% ($\alpha = 0.6$) confidence set can we rule out two-peak functions.
From our confidence sets, the data supports two peaks out of three peaks
in the CMB power spectrum (for $0 \le \ell \le 1100$).

We calculate ranges for the peak heights and locations in Table 1.
Figure 2 shows the non-parametric fit with confidence intervals
for peak heights and locations at 95\% confidence.
Finally, we computed confidence intervals for the ratios of successive peaks
under a three-peak model.
The 95\% confidence interval for the
ratio of the height of the first peak 
to the height of the second peak is
$(1.06,4.27)$.
The 95\% confidence interval for the
ratio of the height of the second peak to
the height of the third peak is
$(0.41,2.5)$.
This rules out equal heights for the first two peaks
at the 95\% level.
These results are consistent with Hu et al. (2001) who find much
stronger constraints on the height-height ratios by fitting to
cosmological models.

\section{Discussion and Conclusions}

We present an application of a new and powerful non-parametric
technique to CMB temperature data.
Past approaches were
based on complicated cosmological models or parameterized forms.
There is superb visual agreement between the non-parametric fit and the 
best fitting cosmological model. Quantitatively, we provide
constraints on the peak locations, heights, and height ratios of
the power spectrum.
These constraints can be used to place corresponding limits on
the cosmological parameters that they describe. For instance, Hu et al. (2001)
derive relationships which could in principle, be used for this purpose.

At the $2\sigma$ confidence level, we find
at least one peak in the current
CMB power spectra data, while at the $1 \sigma$ level,
we find two or more peaks.
Only for a very low confidence, 40\%, can we rule out two peak functions.
Therefore, the data do not yet show the three expected peaks
for $\ell \le 1100$ (in the three CMB datasets examined here).
There are two explanations for this: the model is right,
but there is insufficient precision in the current data,
or the model is wrong. 
If in fact the errors on the current measurements are simply too large,
then these standard errors would have to be reduced to
one-third of their current values
to rule out a two-peak spectrum at the 95\% confidence level.
This suggests a range for the maximum required errors
for future CMB experiments (via MAP and Planck)
to ``discover'' three peaks in the CMB spectrum.

We point out that the lack of assumptions used 
to arrive at our best fit is conservative.
On the other hand, results from fitting assumed cosmological models
are optimistic, since those
models all have a multi-peaked spectra (e.g. Hu et al. 2001).
While the physical underpinnings for cosmological models
are well founded, the last 50 years (or even five) have
seen radical changes in those models which best fit
the data.
Therefore, a method to describe the CMB that is ``cosmology free'' has
scientific value. Finally, we note that the methods described here
can be applied to the many astrophysical problems that are not
well suited for standard parametric techniques.

\vspace{0.2in}
This work was done in collaboration with the Pittsburgh
Computational Astrostatistics Group (\verb+www.picagroup.org+).
The authors would like to thank the referee for suggestions which
improved the readability and usefulness of this work.
RCN, LW, and CG were partially supported by NSF KDI grant DMS-9873442.

\vspace{0.2in}
During the refereeing process of our paper, two related papers came to our
attention
(Durrer, Novosyadlyj \& Apunevych 2001; Douspis \& Ferreira 2001).
These papers perform a model--independent measurement of the CMB power
spectrum but they are not
non-parametric estimates of the CMB acoustic peaks, as discussed herein,
since they use phenomenological models to describe the underlying
power spectrum. It is interesting to note however, that all three analyses
find low statistical significances for the detection of the second and
third peaks. We await higher precision measurements of the CMB power
spectrum to secure the detection, location, shape and amplitude of these
peaks.

\nopagebreak
\begin{deluxetable}{ccc}
\tablenum{1}
\tablewidth{0pt}
\tablecaption{Three Peak Confidence Intervals}
\tablehead{
\colhead{Peak} & \colhead{Location} & \colhead{Height}}
\startdata
1 & (118,300) & (4361,8055) \nl
2 & (377,650) & (1829,4798) \nl
3 & (597,900) & (1829,4688) \nl
\enddata
\end{deluxetable}

\begin{figure}
\epsscale{0.4}
\plotone{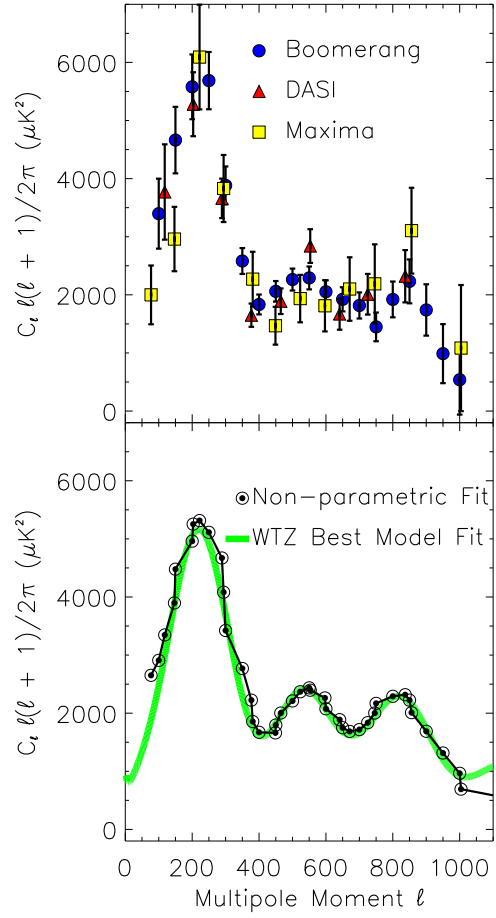}
\caption[]{The {\bf top} panel shows the raw CMB data from the
Boomerang, MAXIMA, and DASI experiments. The {\bf middle} panel is
the raw CMB data over--plotted with
the best fit using the non-parametric technique.
The {\bf bottom} panel shows the non-parametric fit against the best cosmological
model fit from Wang, Tegmark, and Zaldarriaga (2001). }
\end{figure}

\begin{figure}
\plotone{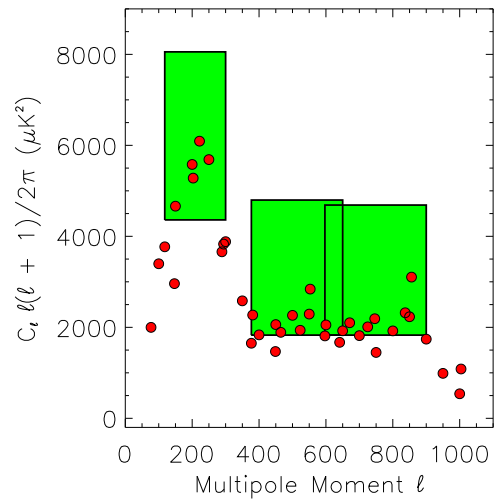}
\caption[]{The ranges on the locations and peak heights of the
non-parametric three peak (and two dip) fit function (95\% confidence).}
\end{figure}

\end{document}